\documentclass[useAMS,usenatbib,usegraphicx]{mn2e}

\usepackage{amsmath}

\def\fermi{{\it Fermi\/}}

\def\cha{{\it Chandra\/}}

\title[A Curious X-ray source in the outskirts of GLIMPSE-C01]{A 
Curious Source of Extended X-ray Emission
in the Outskirts of Globular Cluster GLIMPSE-C01}
\author[N. Mirabal]{N. Mirabal$^{1}$\thanks{E-mail:
mirabal@gae.ucm.es}\\
$^{1}$Ram\'on y Cajal Fellow; Dpto. de F\'isica At\'omica,
Molecular y Nuclear, Universidad Complutense de
Madrid, Spain\\}
\begin{document}

\date{}

\pagerange{\pageref{firstpage}--\pageref{lastpage}} \pubyear{2009}

\maketitle

\label{firstpage}

\begin{abstract}
We report the discovery of an unusual source of extended X-ray emission 
CXOU J184846.3--013040 (`The Stem') located
on the outskirts of the globular cluster GLIMPSE-C01. 
No point-like source falls within the extended emission which has an
X-ray luminosity $L_{X}$(0.3--8 keV) 
$\sim 10^{32}$ ergs s$^{-1}$ and a physical
size of $\sim 0.1$ pc at the inferred distance to the cluster. 
These X-ray properties are
consistent with the pulsar wind nebula (PWN) of an unseen
pulsar located within the 95-percent confidence error 
contour of unidentified \fermi\ $\gamma$-ray 
source 0FGL J1848.6--0138. However, we cannot
exclude an alternative interpretation that postulates X-ray 
emission associated with a bow shock produced from
the interaction of the globular cluster and 
interstellar gas in the Galactic plane. Analysis of the X-ray data
reveals that `The Stem' is most significant in the 2--5 keV band,
which suggests that the emission may be dominated by 
non-thermal bremsstrahlung
from suprathermal electrons at the bow shock.
If the bow shock interpretation is correct, these observations
 would provide compelling evidence that GLIMPSE-C01 
is shedding its
intracluster gas during a galactic passage. Such a direct
detection of gas stripping would help clarify
a crucial step in the evolutionary history of globular clusters. 
Intriguingly, the data may also accommodate a new type of X-ray source.  
\end{abstract}

\begin{keywords}
X-rays: general -- gamma rays: observations -- globular clusters: general
-- stars: neutron
\end{keywords}

\section{Introduction}
For decades, the Galactic plane has been a source of amazement and
unexpected X-ray discoveries
\citep{giacconi,giacconi2,tanaka,voges}. 
The current
generation of X-ray satellites \cha\ and {\it XMM-Newton} 
are the latest to
extend the census of X-ray sources along
the Galactic plane to unprecedented depths \citep{motch,champlane}.
Together these experiments have revealed a rich tapestry of 
X-ray binaries, 
neutron stars, pulsar wind nebulae (PWNe), 
supernova remnants, coronal emitting stars,
and non-thermal filaments along the Galactic plane. However, 
for every physical object identified with high confidence, 
the literature is plagued by an equal or
greater amount of X-ray sources that
continue to elude firm identifications
\citep{wang,muno}. 
Frustratingly, a combination of high 
extinction, excessive crowding, and the lack of direct distance 
indicators prevent further progress in the identification of many of these
sources.

During a routine multiwavelength 
survey of $\gamma$-ray error contours produced by the \fermi\ mission
\citep{mirabal}, we
have encountered the latest puzzling source 
connected with the Galactic plane. 
The object CXOU J184846.3--013040 -- which we nickname `The Stem' -- 
corresponds to
 a source of extended emission in the outskirts of  
GLIMPSE-C01. The low-latitude Globular Cluster
GLIMPSE-C01 is located in the Galactic
plane at $(\ell, b)=(31.\!^{\circ}3,-0.\!{^\circ}1)$ and was discovered during 
Galactic plane scans conducted by the {\it Spitzer Space Observatory} 
\citep{kobul} and the Two Micron All Sky Survey  \citep[2MASS;][]{simpson}. 
Subsequently, superb observations with the {\it Chandra X-ray Observatory} 
revealed a population of X-ray point-like sources with luminosities
$L_{X} > 6 \times 10^{31}$ ergs s$^{-1}$ around the core of the cluster 
\citep{pooley}. 

\begin{figure*}
\hfil
\includegraphics[width=2.1in,angle=0.]{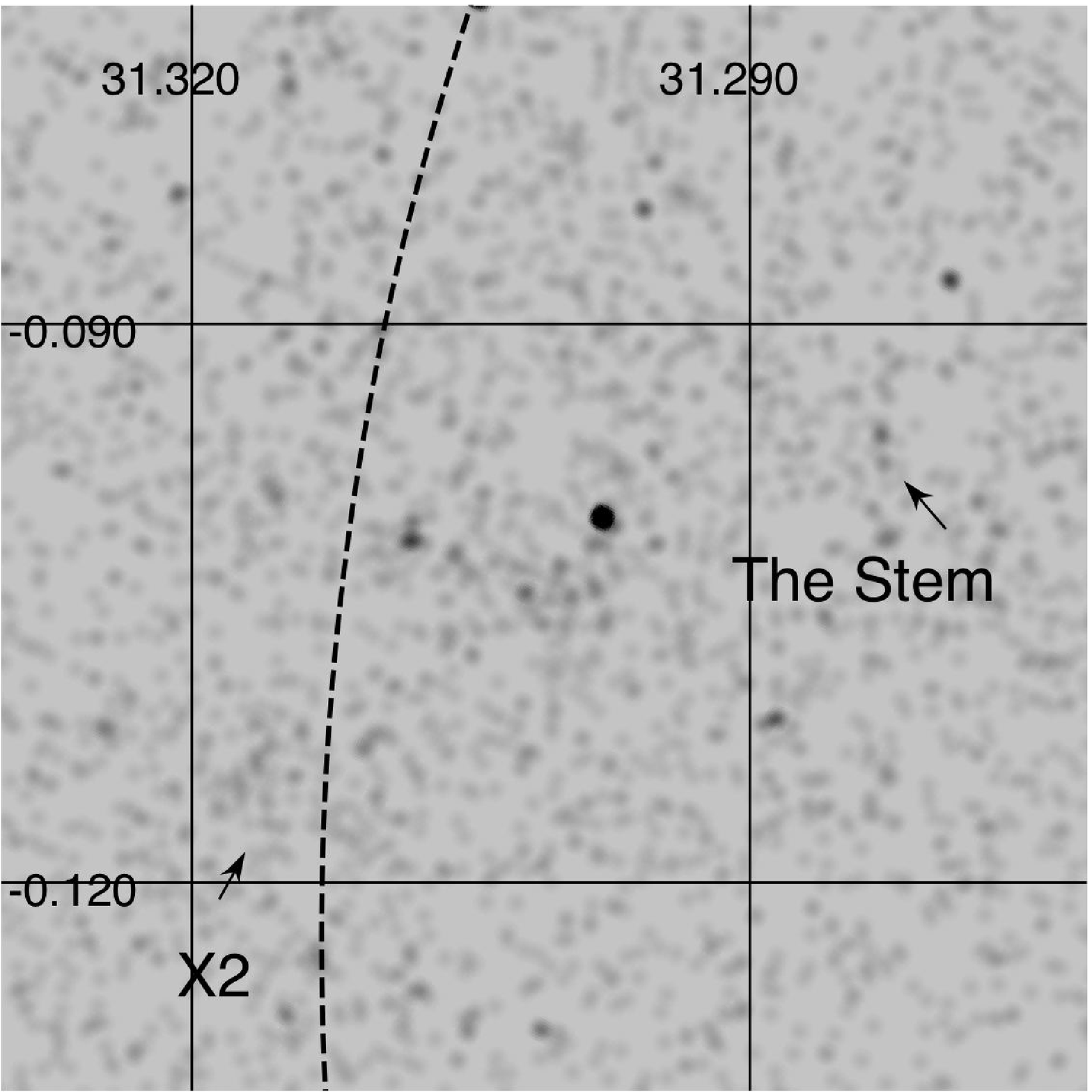}
\includegraphics[width=2.1in,angle=0.]{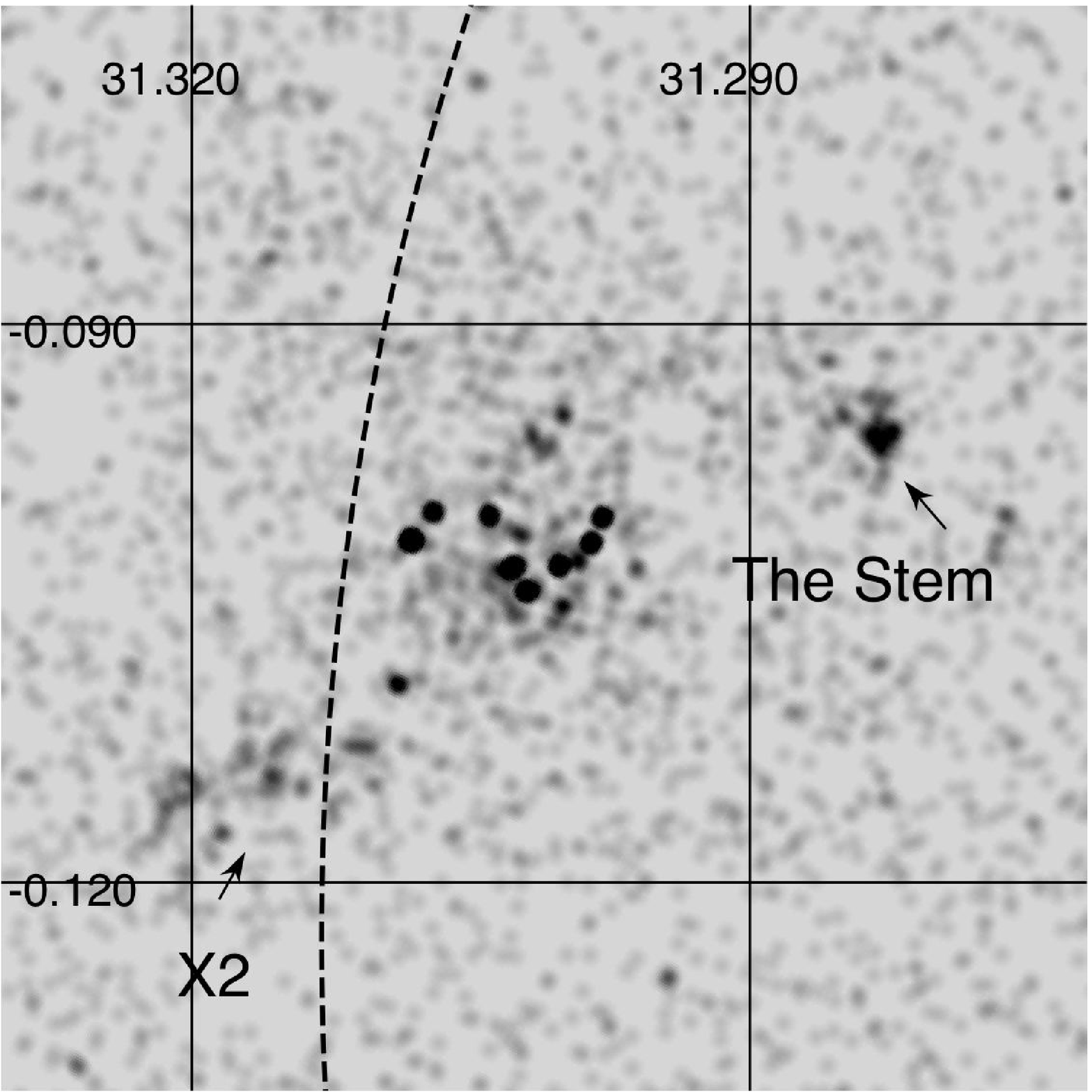}
\includegraphics[width=2.1in,angle=0.]{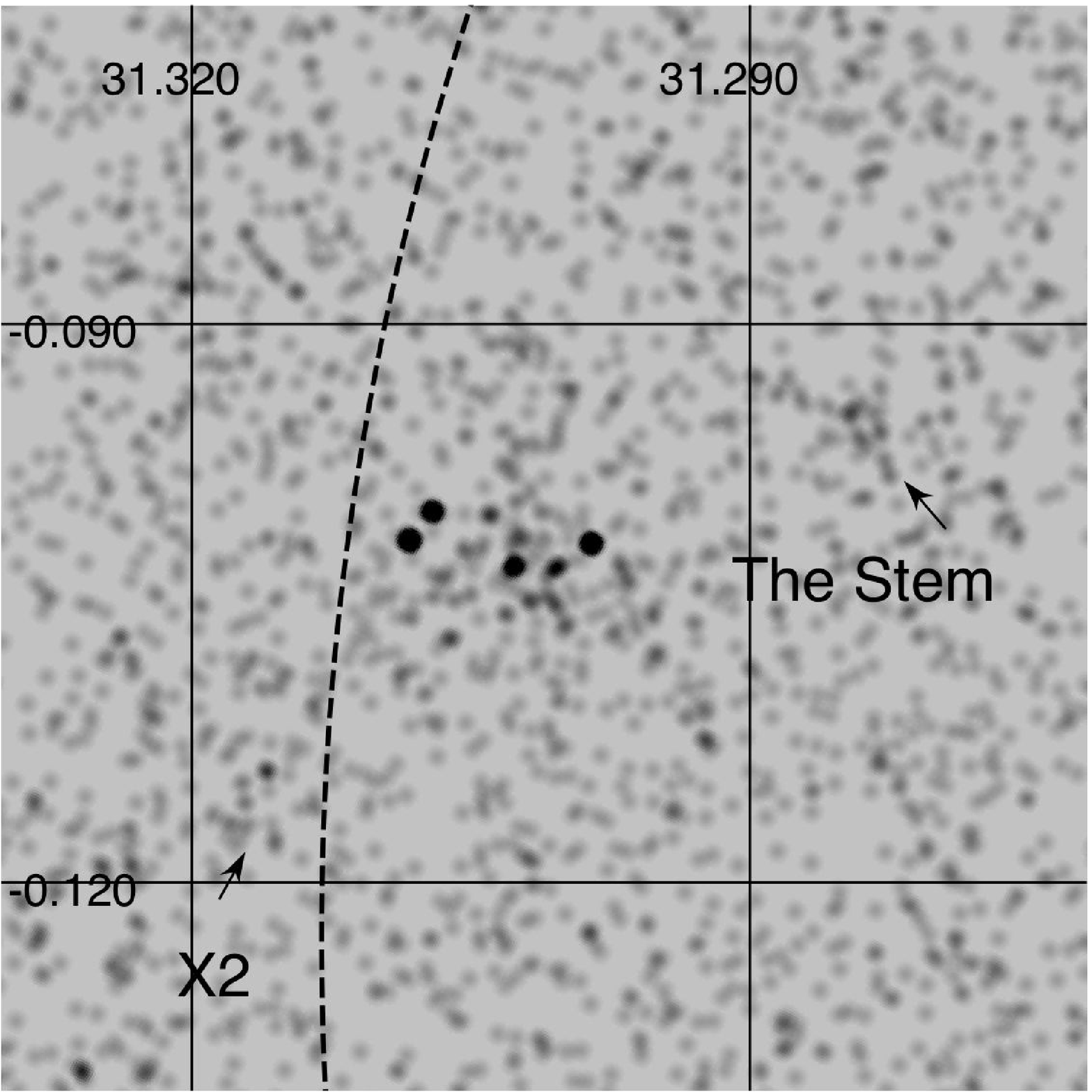}
\hfil
\caption{{\it Left panel:} 0.3--2 keV \cha\ image.
{\it Middle panel:}  2--5 keV \cha\ image. {\it Right panel:}  5--8 
keV \cha\ image overlain with the \fermi\ error
contour ({\it dashed line}) derived by \citet{abdo1}.  
The X-ray images have been smoothed with a
Gaussian kernel with radius $r_{k}$ = 2.5\arcsec.
The superposed arrows mark the location of the `The Stem'; 
as well as the  extended trail emission
designated X2.  The field size is $\sim$ 3.5\arcmin $\times$ 3.5\arcmin shown 
in Galactic coordinates. North Galactic Pole is in the direction of the
top of the figure and longitude increases toward the left. 
}
\label{figure1}
\end{figure*}

`The Stem' is an extended source of emission placed well 
outside the cluster centre 
and is remarkable for two completely different reasons. First, it 
lies within the 95-percent confidence error contour of
unidentified \fermi\ source 0FGL J1848.6-0138 \citep{abdo1,luque}. 
If powered by a compact object, `The Stem' would 
constitute a formidable counterpart candidate 
for the $\gamma$-ray emitter.
The second reason, unrelated to the $\gamma$-ray emission,
is the fact that the position of 
GLIMPSE-C01 along the Galactic plane indulges us in 
that rare geometrical arrangement where a Globular cluster is 
caught in the act of crossing the Galactic plane. This is
the precise moment in which the cluster is believed to interact
with the interstellar medium (ISM) and proceed to shed some of the 
intracluster medium accumulated from the continuous 
mass loss of individual stars within the cluster \citep{frank,faulkner}.

In order to weigh the merits of each possibility, we
analyse archival \cha\ images; as well as infrared observations
of `The Stem' obtained with the {\it Spitzer Space Telescope}. 
The organization of the paper is as follows. \S 2  describes
the observations. Spectral fits are summarized in \S 3.
In \S 4 we discuss alternative models for the X-ray emission. 
Finally, conclusions and future work are presented in \S 5.

\section{Observations}

X-ray observations of GLIMPSE-C01 and surrounding regions were 
obtained with
the Advanced CCD Imaging Spectrometer (ACIS) onboard
the \cha\ X-ray Observatory on 2006 August 15--16 UT. The
total exposure time for the observation was $\sim$ 46 ks.
The \cha\ data were analysed using CIAO version 4.1.1 and
version 4 of the calibration database (CALDB).
Apart from the population of point-like sources associated with the cluster, 
the ACIS image
reveals an extended source of emission CXOU J184846.3--013040 centered 
at (J2000.0) R.A.=$18^{\rm h}48^{\rm m}46.\!^{\rm s}3$,
decl.=$-01^{\circ}30\arcmin40\arcsec$.  The source lies 79 arcsec from
the cluster centre and appears
to have an extent of  $\approx$
5 arcsec in radius, although
it seems slightly elongated. We designate
this source  `The Stem' that in this context can be  
understood as   
the front part of a moving object (GLIMPSE-C01), as will be clear later on. 

Figure~\ref{figure1} shows the resulting smoothed \cha\ ACIS-S3
chip divided in soft (0.3--2 keV), medium (2--5 keV), and hard (5--8 keV)
bands.  In all cases, the X-ray images have been smoothed with a 
Gaussian kernel with radius $r_{k}$ = 2.5\arcsec.
We find that the \cha\ images are dominated by point-like sources surrounding
the cluster core already discussed by \citet{pooley}. Away from the cluster
core, two 
separate regions of extended emission are visible. `The Stem' corresponds to 
the area located on the northwest corner. There is also evidence for 
a fainter trail of extended emission (which we dub X2) that 
can be made out in the southeast corner of GLIMPSE-C01. 
Both `The Stem' and X2 are revealed most prominently 
(9.4 $\sigma$ level of significance for `The Stem') 
in the medium (2--5 keV) band. For completeness, we note that 
the said sources do not show extreme variability over the span of
the observations.

\begin{figure}[t]
\hfil
\includegraphics[width=3.1in,angle=0.]{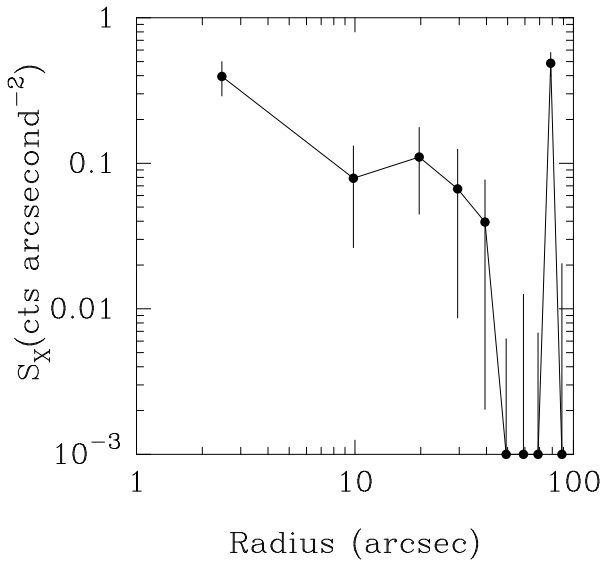}
\hfil
\caption{Radial brightness profile $S_{X}$ 
in the 2--5 keV band as a function of radius. 
The bright jump in the profile 79 arcsec from the centre
corresponds to the location of the extended source of emission (`The Stem').
}
\label{figure2}
\end{figure}

To investigate the
radial distribution of the extended X-ray emission $S_{X}$ across the cluster, 
we removed all the bright point-like sources within the 
half-light radius of the cluster. We next proceeded to  extract 
net counts in a sequence of tangent circles (of
radius equivalent to 10 pixels)
starting at the centre of the globular cluster. Background counts
were computed from source-free regions located $\sim 2.5\arcmin$ from
the cluster center. 
To determine $S_{X}$, the 
number of counts in each region were divided by its respective area in 
square arcsec. The derived radial brightness profile
is shown in Figure~\ref{figure2}. 
The profile shows an initial peak 
dominated by extended emission associated with the globular cluster 
\citep{pooley}. Away from the centre, the most striking feature in the 
surface brightness profile is the jump that occurs near 79 arcsec
that coincides with the position 
of `The Stem'. Such a sharp discontinuity in the profile  
excludes significant contamination from background emission in 
surrounding regions.

\section{Spectral Analysis}
In order to characterize the X-ray spectrum of `The Stem', 
counts were extracted from a circular region with a 20-pixel
radius (9.8\arcsec)  centered 
at (J2000.0) R.A.=$18^{\rm h}48^{\rm m}46.\!^{\rm s}3$,
decl.=$-01^{\circ}30\arcmin40\arcsec$. The background was
extracted from a source-free region with similar radius.
Within this region, we obtained 103 source events in the 0.3--8 keV
band. As a last step, the extracted photons were 
grouped to a minimum of 15 counts per bin. Spectral 
fits to X2 are omitted since 
it comprises a series of low-significance structures
rather a single continuous region.
Herein, we shall discuss X2 only in terms of specific X-ray bands. 

\begin{figure}
\centerline{
\hfil
\includegraphics[width=0.95\linewidth]{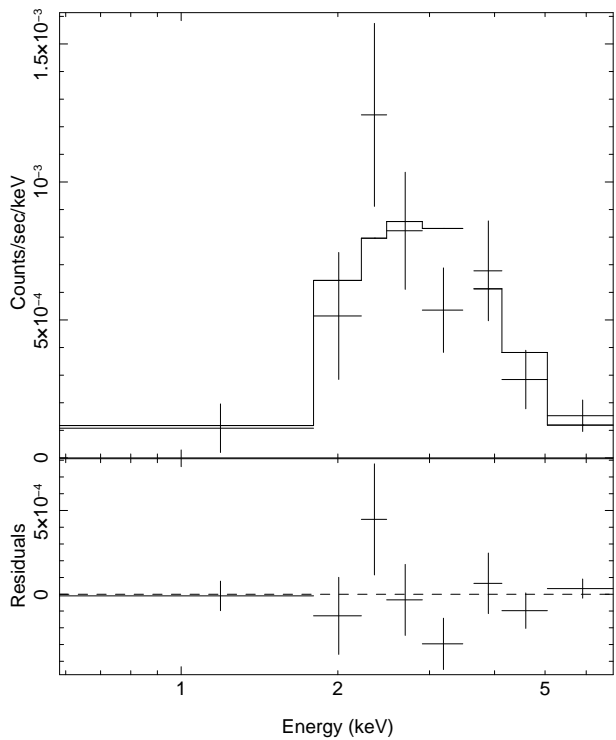}\hfill
\hfil
}
\caption{\cha\ ACIS-S3 spectrum of `The Stem'
and best-fitting absorbed power-law model as described in the text.
}
\label{figure3}
\end{figure}

The resulting spectrum was 
modelled using the X-ray fitting
package XSPEC. Several models provide statistically acceptable
fits to the data. A power-law model results in a steep photon index
$\Gamma = 3.1 \pm 0.8$ and a rather high Galactic H I
column density $N_{\rm H}$ = 
$(6.0 \pm 2.6) \times 10^{22}$ cm$^{-2}$ ($\chi^{2}_{red} = 1.1$). Here
and throughout the text, error on individual parameters are
quoted at the one-sigma level.
The derived value of $N_{\rm H}$ is in excess of the 
total Galactic column density $N_{\rm H}$ = 
$1.7 \times 10^{22}$ cm$^{-2}$ obtained from the 
nH tool\footnote{http://heasarc.gsfc.nasa.gov/cgi-bin/Tools/w3nh/w3nh.pl}.
The fit worsens if the absorption if $N_{\rm H}$ is treated as a fixed
parameter. Because of the low number of photons associated with 
other point sources around the core, 
no additional constraints can be placed on the column density 
for the cluster. 
The resulting spectrum and best-fitting power-law 
model for `The Stem' are shown in Figure~\ref{figure3}.

For comparison, a black-body model yields $k$T = $0.9 \pm 0.2$ keV
and a column density $N_{\rm H}$ =
$(3.0 \pm 1.5) \times 10^{22}$ cm$^{-2}$ ($\chi^{2}_{red} = 1.5$)
closer to the value inferred using the nH tool. A thermal 
bremsstrahlung model with $k$T = $3.0 \pm 1.2$ keV and
$N_{\rm H}$ = 
$(4.6 \pm 1.2) \times 10^{22}$ cm$^{-2}$ ($\chi^{2}_{red} = 1.3$) 
is also a possibility. Lastly, a 
Raymond-Smith model with solar abundance results in 
$k$T = $2.1 \pm 0.7$ keV and
$N_{\rm H}$ =
$(5.5 \pm 1.4) \times 10^{22}$ cm$^{-2}$ ($\chi^{2}_{red} = 1.0$).

With the variety of models allowed by the observation, 
attempting multi-component models
become a superfluous exercise. 
Instead, we choose to restrict the rest of our analysis 
to the models considered here.
It might be the case that the actual X-ray spectrum is best explained by 
multi-component fits, but better and deeper observations will be needed
to settle this point.  Independent of the specific model, 
we derive an unabsorbed flux in the
0.3--8 keV band of $\approx 6.3 \times 10^{-14}$
erg cm$^{-2}$ s$^{-1}$ for a fixed column 
density $N_{\rm H}$ =
$1.7 \times 10^{22}$ cm$^{-2}$.

\section{Possible Interpretations of the Extended Emission}

\subsection{A Pulsar Wind Nebula within the error contour of 0FGL 
J1848.6--0138?}
\begin{figure}
\centerline{
\hfil
\includegraphics[width=0.95\linewidth]{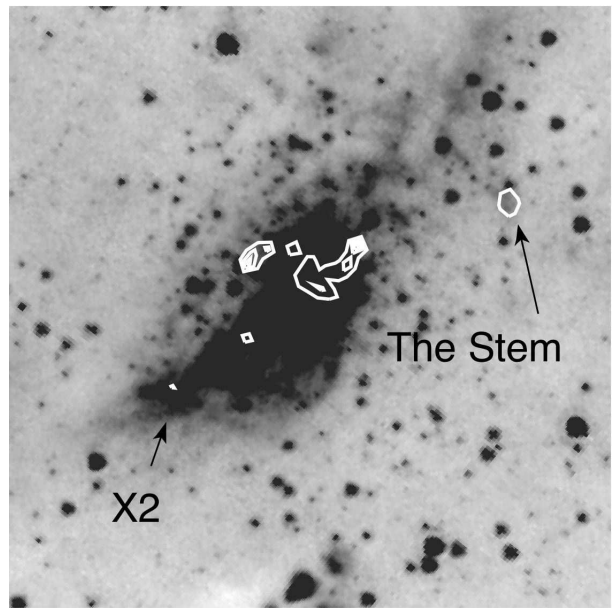}\hfill
\hfil
}
\caption{Infrared composite constructed from IRAC 3.6, 5.8, 
and 8.0 $\mu$m images obtained by the {\it Spitzer
Space Telescope}. The overlain contours indicate the brightest
X-ray features. The position
of `The Stem' is clearly void of bright infrared point-like sources.
Also shown is the position of low-level emission X2. 
Field is plotted in Galactic coordinates (latitude increases upward) 
and it spans 3.5\arcmin $\times$ 3.5\arcmin.}
\label{figure4}
\end{figure}

A PWN seems to offer a natural explanation for
`The Stem'. In this picture,
the pulsar wind shocks with the ISM 
and creates a nebula morphology that emits synchrotron radiation 
\citep{gaensler}.  In order to evaluate this 
possibility further, we examined an infrared map composed from  3.6,
5.8, and 8.0$\mu$m images obtained by the 
{\it Spitzer Space Telescope} with the IRAC instrument on 2004 April 21
\citep{kobul}.  Figure~\ref{figure4} shows the three-color image of this region
overlain by the X-ray contours derived from the \cha\ observation. 
Note that the area around `The Stem' itself 
is void of bright stars, which appears to exclude 
a chance alignment of coronal emitting stars or novae that could explain 
the extended X-ray emission \citep{anderson}. 

\begin{figure}
\hfil
\includegraphics[width=0.95\linewidth]{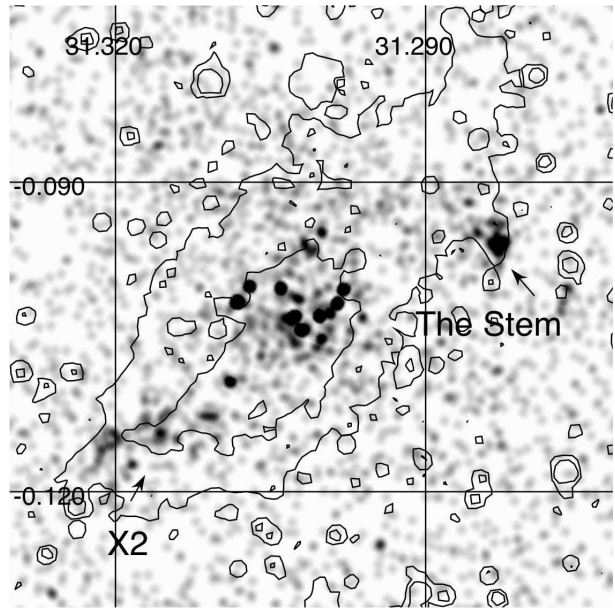}
\hfil
\caption{
Smoothed Chandra observation of GLIMPSE-C01 in the 2-5 keV energy band, showing
`The Stem' and X2. Overlain contours correspond to the infrared
emission constructed from an IRAC 8.0 $\mu$m image obtained with the Spitzer
Space Telescope.  The figure is in
Galactic coordinates (North Galactic Pole is up).
The field size is $\sim$ 3.5\arcmin $\times$ 3.5\arcmin.
}
\label{figure5}
\end{figure}

Turning the bands around, Figure~\ref{figure5} 
shows the \cha\ X-ray image overlaid
with the infrared contours. From the figure, we can see that `The Stem' is 
properly aligned with a distinct infrared structure growing 
continuously from the main plume of infrared emission. 
We argue that the morphological correspondence
suggests that the extended X-ray emission lies at the same distance as
GLIMPSE-C01. As noted by \citet{pfahl}, 
a fraction of neutron stars or potential neutron star progenitors 
could be ejected from the core but remain bound to the cluster 
as a result of dynamical interactions. 
The placement of `The Stem' may be the result of
one such interactions. However, actual physical
membership in the cluster is not a requirement for a PWN interpretation.

Assuming that indeed `The Stem' lies at the distance to
the globular cluster $D = 4$ kpc \citep{pooley}, the unabsorbed 
flux implies an X-ray luminosity $L_{X} \sim 10^{32}$ ergs s$^{-1}$ 
in the 0.5--8 keV
band. Such value is generally consistent with observed values
for well-studied PWNe \citep{cheng}. Similarly, the derived spatial 
extent of the source ($\approx$
5 arcsec in radius) translates to $\sim$ 0.1 pc at this distance. 
We note that this value is  also perfectly 
in line with the values reported for the termination radii of 
PWNe (Cheng et al. 2004). The overall agreement with typical values
of known pulsars provides encouraging evidence for a PWN
interpretation. 

Owing to the lack of radio
emission, \citet{kobul} have also reported
a radio upper limit of 1.4 mJy for point-like sources within this region
based on 1.4 GHz observations obtained with the Very Large Array (VLA).
An integration of the radio upper limit
from $10^{7}$ to $10^{11}$ Hz with an assumed flux density  
$S_{\nu} \propto \nu^{\alpha}$ and spectral index of $-$0.4
implies a
a radio luminosity upper limit $L_{R} < 10^{30}$ ergs s$^{-1}$. 
We note that such low value
is not unprecedented for the emission of PWNe in radio \citep{frail}.

It is only when one turns to the spectral properties of the source that
a potential problem arises. A pure synchrotron model, 
the spectrum should 
be characterized by a power-law spectrum with photon indices 
$2 < \Gamma < 2.5$ \citep{gaensler}. 
The power-law fit of the observed spectrum from `The Stem' gives
$\Gamma \sim 3.0$ consistent with a rather steep electron spectral
index $p > 4.0$. As a 
possible way out of this challenge, we postulate that the observed
X-ray spectrum most likely comprises a complicated mixture of non-thermal
and thermal components of the as-yet unresolved X-ray point-like source
powering the nebula. We point out that a two-component model consisting 
of a blackbody plus a power-law component requires 
a non-thermal to thermal
flux ratio (0.3--8 keV) in the range of $\approx 2$. Such ratio is consistent
with that observed for rotation-powered pulsars \citep{cheng}.

Previous X-ray observations of PWNe have also shown that the pulsar
itself is usually revealed as a point source within the nebular morphology
\citep{gaensler}. In this instance, 
it is possible that the pulsar itself is disguised by an overluminous
clump of extended emission. Otherwise, the pulsar might have moved well outside
the observed pulsar wind nebula. Moving with a transverse velocity $\sim$
500 km s$^{-1}$, it would require the pulsar at least $10^{4}$ yr to reach
the edge and even longer to disappear from the 
vicinity of the PWN. As stated earlier, 
deeper X-ray observations are needed to clarify these alternatives.

On the basis of a PWN interpretation, 
`The Stem' immediately becomes a potential counterpart candidate
of the unidentified \fermi\ $\gamma$-ray
source 0FGL J1848.6-0138 \citep{abdo1} and more loosely
HESS J1848--018 \citep{chaves}. 
A previous multiwavelength search of the error
\fermi\ 95-percent confidence error contour of 0FGL J1848.6-0138
failed to produce any
prominent counterpart candidate of the unidentified
$\gamma$-ray source \citep{luque}. Considering that a remarkable
number of pulsar PWNe have been found to be coincident
with both EGRET and \fermi\ error contours \citep{mallory,abdo1},
it is conceivable that `The Stem' is powering the $\gamma$-ray emission. 
Provided that the ratio of X-ray luminosity $L_{X}$ to spin-down power 
\.{E} lies in 
the range $10^{-4}$--$0.1$ \citep{halpern}, the hypothesized 
spin-down power of `The Stem' pulsar should be 
$10^{33}$--$10^{37}$ ergs s$^{-1}$. These values of 
\.{E} are consistent with
the range derived among the
 $\gamma$-ray pulsars discovered by \fermi\ \citep{abdo2}. 

Though this scenario appears feasible, it is important to realize that 
`The Stem' is located at the tangent point
of the Crux-Scutum spiral arm where there is a high density
of unusual sources \citep{vallee}. As a result, we warn that it 
will be difficult to prove
a direct association with any $\gamma$-ray source in this region.  

\subsection{A globular cluster being stripped by  the Galactic plane?}
Possibly the greatest challenge to the preceding 
explanation is the odd placement of `The Stem' 
in the outskirts of the globular
cluster GLIMPSE-C01. A typical globular cluster is expected
to accumulate $10^{2}$ -- $10^{3}$ $M_\odot$ of gas mainly due to 
mass-loss from  red giant and asymptotic giant branch stars 
within the cluster \citep{rood}. Despite dedicated efforts in
mapping globular clusters, 
 an actual detection of this hypothesized intracluster 
gas remains elusive \citep{vanloon}.
In order to explain the observed discrepancy,
it has been suggested
that most of the hypothesized intracluster gas could be 
shed during globular cluster passages
through the Galactic plane \citep{frank,faulkner}. 

Since globular clusters are predicted to be moving at velocities
$\sim$ 100--300 km s$^{-1}$, a 
particular associated prediction is that a bow shock should form 
at the interaction of the intracluster gas 
and the ISM. Yet, X-ray imaging 
have failed to find definite observational evidence for
such interactions \citep{krock,hopwood,okada}.
We argue that previous attempts to locate bow shocks have failed 
mainly due to two effects: (1) the majority of globular clusters reside
within the low-density Galactic halo and (2)  
the waiting time for a future Galactic plane 
passage ($\sim 10^{8}$ yr) is largely unpractical relative to a typical 
human lifetime.

At a low Galactic latitude $b = -0.\!^\circ1$, GLIMPSE-C01
is ideally placed to search for such an encounter. Interestingly, 
`The Stem' coincides nicely with a continuous
infrared morphology growing out of the main infrared plume
(see Figure~\ref{figure5}). 
In fact, \citet{kobul} originally
suggested that much of the 
extension of GLIMPSE-01 both in infrared and submillimeter
could be connected with intracluster debris
stripped by the Galactic ISM. However, these
authors lacked the
X-ray imaging that could directly trace the bow shock from such
interaction.  

When a moving
object interacts supersonically 
with the ambient medium, one would expect the formation 
of a nebula with cometary morphology including a `head' and a `fan-like tail' 
\citep{olbert}. In our case, the X-ray image displays an enhanced 
region (`The Stem') that could only correspond the apex of the interaction
(`head'). However, we find no evidence for the expected cometary morphology. 
An ellipse fit to `The Stem' in the \cha\ image
gives an ellipticity of 0.64. For comparison, a fit of the correponsing
infrared contour in the {\it Spitzer} image
indicates an ellipticity of 0.44 nearly aligned with the X-ray shape.
The consistency of the position angles derived in X-rays and infrared 
favors the physical association of
`The Stem' with GLIMPSE-C01. Given the latter consideration, 
deeper observations are needed to confirm or rule out the presence 
of low-level cometary structure surrounding `The Stem'.

If we adopt the bow shock interpretation,
`The Stem' would mark the apex of latest impact (moving south-north
in galactic coordinates). Accepting this hypothesis,  
the stand-off distance of the
bow shock $r_{s}$ can be approximated from the balance between the medium
ram pressure
and the momentum flux of the stellar winds associated with
red giant and asymptotic giant branch stars in the cluster,

\begin{equation}
r_{s} = \left(\frac{\dot M v_{ml}}{4 \pi \rho v_{GC}^{2}}\right)^{1/2} 
\approx 0.6 n_{0}^{-1/2}~pc
\end{equation}

for a stellar mass-loss rate $\dot M$ =
$10^{-5} M_\odot$ yr$^{-1}$ ,
mass-loss velocity $v_{ml}$ = 100 km s$^{-1}$, 
globular-cluster velocity
$v_{GC}$ = 100 km s$^{-1}$, and ISM density $\rho \approx 1.7 \times 
10^{-24} n_{0}$ g cm$^{-3}$ (where $n_{0}$ corresponds
to the ambient ISM density). The derived stand-off distance $r_{s}$ 
is compatible with the observed location of the `The Stem' at $\approx 1.5$
pc from the center of the cluster. In this interpretation, 
the low-level emission regions around X2 could be consistent with
ISM that was previously impacted and 
now forms a trail of shocked gas (Figure~\ref{figure1}).

Next, we need to 
evaluate the energetic budget of the bow shock.
\citet{faulkner} have already 
provided excellent estimates to compute the X-ray luminosity 
that can be emitted in bow shocks around a globular cluster. The 
actual energy input $Q_{in}$ associated with the globular cluster
can written as
 
\begin{equation}
Q_{in} = \dot M \frac{(v_{GC}^{2} + v_{ml}^{2})}{2} \approx 6 
\times 10^{34} ergs~s^{-1}
\end{equation}

using $\dot M$ = 
$10^{-5} M_\odot$ yr$^{-1}$, $v_{ml}$ = 100 km s$^{-1}$, and 
$v_{GC}$ = 100 km s$^{-1}$. With average 
densities $n \approx$ 0.1--1 cm$^{-3}$ along the Galactic
plane, \citet{krock}
estimated that the corresponding fraction of the energy 
input that could be transferred
to X-ray emission lies between 10$^{-4}$ and 0.1 of $Q_{in}$. Converting
$Q_{in}$ implies an X-ray luminosity in the range $6 \times 
10^{30}$  -- $6 \times 10^{33}$ ergs s$^{-1}$
of the total. This range nicely brackets 
the derived X-ray luminosity $L_{X} \sim 10^{32}$ ergs s$^{-1}$
at the distance of GLIMPSE-C01. 

Spectrally, the interpretation is more complicated. Postshock temperatures
$T_{post}$ from a bow shock 
are expected to be $T_{post} \sim 1.4\times 10^{5}$ $v^{2}_{100}$ K, where
$v_{100}$ is the velocity of the cluster in units of 
100 km s$^{-1}$ \citep{krock}. X-ray photons at this temperature 
would be most abundant in the 0.1--0.5 keV band. Instead, the
 X-ray temperature  derived from
Raymond-Smith fit \citep{raymond} to the `The Stem'  indicates 
much harder emission with $T \sim 2 \times 10^{7} K$. The same applies
to the X2 region that is
most prominent in the the 2--5 keV band. Assuming that the X-ray emission
is  thermal in origin,
such high temperature would require a globular  cluster moving at an
unrealistic  $v_{GC} >$ 1,000 km s$^{-1}$.  

\citet{krock} argued that
the hard X-ray emission may instead be formed by non-thermal emission 
generated 
as relativistic electrons in the bow shock inverse Compton scatter
cluster photons to harder X-ray energies. Coincidentally, a fraction 
of the unresolved emission detected 
near the centre is generally hard
(see Figure~\ref{figure2}). This supply of  
`hard' photons from the cluster core could potentially 
scatter off electrons accelerated
within the shock. However, it is not entirely clear that  
there are enough mildly relativistic electrons within the cluster 
to support this process \citep{krock}.

Alternatively, the hard emission could be due to
non-thermal bremsstrahlung produced by a population of
suprathermal electrons at the bow shock \citep{okada}.  
Measurements near the Earth's bow shock have revealed such population
of suprathermal electrons in the 1--20 keV range with 
a power-law spectrum of index $\Gamma > 3$ \citep{gosling}. 
The biggest uncertainty in this scenario
is the actual 
number of suprathermal electrons carried by the bow shock. In order 
to estimate the required value, 
\citet{okada} obtained an expression that relates the 
number of suprathermal electrons in the bow shock to X-ray luminosity
given by
$L_{X} (0.5-4.5 keV) = 7.4 \times 
10^{30} N_{ste} n_{0}$ ergs s$^{-1}$, where $N_{ste}$ denotes the
total 
number of suprathermal electrons in units of $10^{54}$.  
This estimate implies that
a population of $7 \times 10^{54}$ suprathermal electrons must be available 
across the bow shock to reproduce
the observed emission. 

Assuming that the bow shock covers 
10 percent of a sphere of 9 $r_{c}$ radius (where $r_{c} \sim 0.17$ pc 
represents the core radius) with an annulus width of 0.2 pc, we find a 
maximum of suprathermal electrons given by $N_{ste}$  
$\sim 2 \times 10^{55} n_{0} \eta$. Here, $\eta$ represents the fraction
of suprathermal electrons associated with the bow shock. Using the
estimate by \citet{okada}, we find that 
a value of $n_{0} \eta > 0.35$ is required.  As 
previously stated,  it is difficult to estimate $n_{0}$ and $\eta$ 
directly from the data. Therefore, 
future X-ray observations and modelling must clarify this issue.

We close by noting that 
numerous models have been investigated to explain the evacuation of gas
from globular clusters \citep{spergel,freire}. 
However, if the bow shock interpretation of `The Stem' is correct, 
these observations would provide direct evidence 
of a mechanism for removing intracluster
gas on Galactic scales. 
Under this scenario, the vast amount of stellar ejecta accumulated 
during the orbit of the globular cluster would be 
stripped during successive  passages through the
ISM in the Galactic plane.

\subsection{Alternative explanations}
Because we cannot directly derive a distance to the source of extended
emission, there is always a chance that `The Stem' 
is produced by a foreground or background source completely unrelated
to a PWN or a globular cluster passage. Spectral fitting
with an absorbed MEKAL component \citep{mewe}, commonly used
to describe gas in galaxy clusters, results in an unacceptable 
$\chi^{2}_{red} = 2.1$. Furthermore,
the positional coincidence between the X-ray and infrared images, 
alongside the
high extinction estimated 
for this line of sight  $A_{K} \approx 9$ \citep{kobul}
most likely rules
out an extragalactic origin for the X-ray emission. 

Probably, the only
outstanding explanation is a non-thermal radio filament \citep{yusef}. 
However, we find such possibility unlikely as no 
prominent radio-emitting filament has been revealed in
this general region. 
Nevertheless, since none of the proposed explanations has been proven
conclusively, it is important to leave open the option of a new
type of X-ray source with unique properties. Perhaps a novel
emitter remains to be found, but such issue is beyond the
scope of this work.

\section{Conclusions and Future Work}
In view of our discussion, two leading explanations emerge to
explain the source of extended emission in the outskirts of 
GLIMPSE-CO1. First, based purely on its purported 
X-ray luminosity and physical size,  
`The Stem' may represent a PWN lacking a central point-like
source. If so, `The Stem' 
becomes the most notable object
in the 95-percent confidence error contour of unidentified \fermi\ source 0FGL
J1848.6-0138. The alternative model, a potentially more
intriguing explanation for the extended source of emission, is that
`The Stem' traces back to a bow shock
produced as the globular cluster GLIMPSE-C01 passes through the 
Galactic plane. If confirmed, the latter 
explanation would indicate that GLIMPSE-C01 is losing part of 
its intracluster gas in the process. More generally, it would 
provide 
evidence of the systematic stripping of
intracluster
gas on Galactic scales long predicted by a number of 
theoretical models.
However, given the unusual nature of
this source, it is also possible that `The Stem' represents
a new type of X-ray emitter.

With these findings at hand, it is critical to determine the radial velocity
and proper motion of 
GLIMPSE-C01 through alternative means. In addition, 
deeper X-ray observations of
this region are needed to refine the current modelling of the X-ray 
emission. We envision two vital tests
that may conclusively trim these models.  
First, the orbit of the GLIMPSE-C01 must agree with
the placement of the purported bow shock. Second, radio/X-ray pulsation
searches must aim for a pulsar within the extended emission 
and place stricter constraints on the radio component. 
Given the heavy crowding around this region, we believe that
GLIMPSE-C01 makes an exquisite target for the
newly refurbished {\it Hubble Space Telescope} that would result in
the most accurate proper
motion measurements of the globular cluster.

\section*{Acknowledgments}
I thank Fernando Camilo, Eric Gotthelf, Jules Halpern and Daniel Nieto 
for useful comments. I am also grateful to the referee for suggestions
that helped to improve the article. 
I acknowledge support from the Spanish Ministry of Science
and Technology through a Ram\'on y Cajal fellowship.

\label{lastpage}
\end{document}